\documentclass[12pt,preprint]{aastex}
\usepackage{emulateapj5}
\usepackage{psfig}
\shortauthors{E. Vesperini, S. E. Zepf}
\shorttitle{Evolution of globular cluster systems}
\submitted{Published in The Astrophysical Journal Letters, 587, L97-L100, 2003 April 20}
\begin{document}
\voffset=-1.8truecm
\title{Effects of the dissolution of low-concentration globular
clusters on the evolution of globular cluster systems}
\author{E.Vesperini\altaffilmark{1}, S.E. Zepf\altaffilmark{1}}
\altaffiltext{1}{Department of Physics and Astronomy, Michigan State
University, East Lansing, MI, 48824; e-mail:vesperin, zepf @pa.msu.edu}
\begin{abstract}
We investigate the role of dissolution of low-concentration clusters due to
mass loss through stellar evolution on the evolution of the properties
of globular cluster systems (GCSs) in elliptical galaxies. 
Our simulations show that, for an initial 
mass-concentration relationship based on that inferred from 
Galactic globular clusters,  dissolution of low-concentration clusters
leads to the disruption of a large number of clusters. A 
power-law initial globular cluster system mass function (GCMF) similar to
that observed in young cluster 
systems in merging galaxies is transformed by this dissolution into a
bell-shaped GCMF with a mean mass similar to that of old GCSs for all
the galaxies investigated.  
Two-body relaxation and dynamical friction, which are also included in
our simulations, subsequently lead to an additional significant evolution and
disruption of the population of clusters. As shown previously,
when these processes act on a bell-shaped GCMF with a mean mass similar to that
of old GCS, they do not significantly alter the value of the
mean mass. The final GCMFs
are bell-shaped with similar peaks at different radii within galaxies
and between different galaxies, in agreement with current observations.
\end{abstract}
\section{Introduction}
It has long been realized that globular clusters 
evolve dynamically, and many of the observational effects predicted
by theoretical models of the dynamical evolution of globular
clusters have been observed (see e.g.\ Meylan \& Heggie 1997 for a review).
The dynamical evolution of globular clusters is therefore a 
fundamental component of the comparison of observations and theories 
of globular cluster formation to the observed properties of old 
globular cluster systems. 

Observations of starbursts and mergers have shown that
clusters with the high luminosities and the compact sizes
expected for globular clusters at young ages have formed in these environments
(see e.g. Whitmore 2002 for a recent review and references therein). 
However, the young cluster systems have power-law mass functions
while the old cluster systems have roughly log-normal mass
functions. A power-law mass function is also suggested by
many globular cluster formation models (Harris \& Pudritz 1994,  
Elmegreen \& Efremov 1997, Ashman \& Zepf 2001, Elmegreen 2002). 
Therefore, if the systems of massive, dense young clusters are 
to be specifically associated with globular cluster systems, 
the difference in mass functions needs to be understood.
	
	Many theoretical works have shown that dynamical processes, 
by preferentially disrupting low-mass clusters, can easily transform 
a  power-law initial GCMF into a bell-shaped GCMF 
(e.g. Vesperini 1998 and references therein).
However, in order to support the hypothesis that
GCSs had a power-law initial GCMF, it is necessary to verify that 
evolutionary processes can lead to final GCMF parameters consistent
with observations. In particular,
observations show a small radial variation within individual galaxies 
of the mean mass of clusters (see e.g. Kundu, Zepf \& Ashman 2003 for 
a detailed study of the M87 GCS), and a small galaxy-to-galaxy variation 
of  the mean mass of clusters in galaxies with different masses and sizes
(e.g.\ Harris 2001, Kundu \& Whitmore 2001).
Since the efficiency of evolutionary processes is 
expected to depend both on the properties of the host galaxy and,
within individual galaxies, on the galactocentric distance, the
lack of a significant galaxy-to-galaxy and radial variation of the mean mass is
an interesting question to investigate.

In a number of theoretical investigations the long-term dynamical
evolution of GCSs in elliptical galaxies and in the Milky
Way (Okazaki \& Tosa 1995, Vesperini 1997, 1998, 2001, Murali \&
Weinberg 1997, Baumgardt 1998) starting  with  
a power-law initial GCMF similar to that observed in young cluster systems has
been studied. 
These studies which include the effects of two-body relaxation, tidal
shocks, dynamical friction and mass loss due to stellar evolution and
 a range of different initial conditions for the orbital
distribution of clusters have shown that in
models with power-law initial GCMF similar to that observed in young
cluster systems,  
the radial variation within individual galaxies appears to be larger 
than the observational constraints, the final mean masses
 smaller, and the galaxy-to-galaxy variation of the final mean 
mass larger than suggested by the observations.

	In the context of the Galactic GCS, Fall \& Zhang (2001) 
proposed that, for models with a power-law initial GCMF, the mild
radial variation in the final GCMF might be understood  
if the initial population of globulars spanned a narrow range of
pericentric distances, corresponding to a radial anisotropy strongly increasing
with radius. To test whether a velocity distribution with a strong radial 
anisotropy increasing outwards
is a general answer to the absence of large radial mass gradients 
in the GCMF, Vesperini et al. (2002) studied the dynamical evolution
of the M87 GCS; their models incorporated a broad range of initial 
 conditions, including velocity distributions like that
proposed by Fall \& Zhang for the Galactic GCS.
The advantage to studying the M87 GCS is that
stringent observational constraints on both the GCMF 
and the GCS kinematical properties are available.
Vesperini et al.\ (2002) found that no model of long-term
dynamical evolution could fit both the observed
radial profile of the GCMF and the observed kinematics of
the GCS if a power-law initial GCMF was adopted. Models with strong
radial anisotropy at large radii were inconsistent with the
kinematics, while isotropic models produced radial mass
gradients larger than observed.

Whether dynamical processes can turn a
power-law into a bell-shaped GCMF with a final mean mass consistent
with observations while satisfying, at the same time, all the observational
constraints on other GCS properties is therefore still an open question.  

It has been shown in a number of 
theoretical investigations (see e.g. Chernoff, Kochanek \& Shapiro
1986, Chernoff \& Shapiro 1987,  Chernoff \& Weinberg 1990, Fukushige
\& Heggie 1995, Takahashi \& Portegies Zwart 2000 )
that if a star cluster has initially a low concentration and/or if the
stellar IMF contains a sufficient number of massive stars which rapidly evolve
leading to a substantial early mass loss, the cluster response
to this mass loss is an expansion ending  with its complete
dissolution. Although mass loss due to stellar evolution is
included in most theoretical studies of GCSs evolution, it is always implicitly
assumed that individual clusters have an initial concentration large
enough to avoid  dissolution due to mass loss of massive
stars.  The goal of this Letter is to study the role of
this process on the evolution of the properties of globular cluster
systems in elliptical galaxies and, in particular, to explore its
effect on the evolution of the GCMF.

\section{Method and Initial Conditions}
In order to determine the fate of individual clusters we use a method
in which the time evolution of clusters is assumed to occur along a
sequence of King models with evolving mass, size and concentration. 
This method, which was introduced by King 
(1966) and adopted by Prata (1971), Chernoff, Kochanek \& Shapiro
(1986), Chernoff \& Shapiro (1987), Vesperini (1997) to
study the evolution of the Galactic GCS, allows to follow the
evolution of the concentration of individual clusters and, in
particular, to determine whether a cluster evolves toward lower
concentrations and eventually dissolves in response to the initial
expansion triggered by mass loss due to stellar evolution (we consider
a cluster dissolved when its structure is equal to that of a King model
with central dimensionless potential, $W_0$, smaller than
$10^{-2}$; see, e.g., Chernoff \& Shapiro 1987 for a more detailed 
description of the method adopted). This
additional route to disruption is missing in all the previous
investigations studying the evolution of the GCMF.

If its initial concentration is large enough, a cluster  survives
the initial expansion caused by stellar evolution but it loses mass
and it can be  
disrupted because of complete evaporation due to mass loss caused by
internal relaxation or because of dynamical friction.
To follow the time evolution of
the mass of clusters, we use the results of the N-body
simulations carried out by Baumgardt \& Makino (2003) (similar
results are obtained if the results of N-body simulations by Vesperini
\& Heggie (1997) are used).

In each simulation carried out we have studied the evolution of a GCS
initially made of 400000 clusters with initial masses between $10^4~M_{\odot}$
and $10^7 M_{\odot}$  distributed
according to a power-law  GCMF with index 1.8, with galactocentric
distances ranging from 0.16$R_e$ to 20$R_e$ where $R_e$ is the
effective radius of the host galaxy. 
The GCS initial number density distribution
is taken equal to a Navarro, Frenk \& White (1996) profile with scale
radius $r_s\simeq 0.02 R_e$; this is essentially equivalent to a single
power-law with index equal to 3.
An isotropic initial GCS velocity distribution is adopted.

For the stellar IMF of individual clusters we adopt the function suggested
by a recent observational analysis by Kroupa (2001): this is a
two-slope power-law function with index equal to 2.3 for
$m_T<m/m_{\odot}<15$ and index equal to 1.3 for $0.1<m/m_{\odot}<m_T$
with $m_T=0.5~m_{\odot}$. In order to study the dependency of our
results on the value of $m_T$ we have also carried out a set of
simulations for $m_T=0.9~m_{\odot}$. Mass loss due to stellar
evolution is modeled as in Chernoff \& Weinberg (1990).

The initial cluster concentration (the standard concentration, $c$,
used to characterize King models 
and defined as the logarithm of the ratio of the tidal to the King core
radius is adopted here) is assumed to increase
with the mass of clusters as observed for globular clusters in the Milky 
Way (see e.g. Chernoff \& Djorgovski 1989, Djorgovski \& Meylan
1994). As discussed in McLaughlin (2000), the observed trend between mass
and concentration is likely to be primordial. 
For our simulations we have adopted $c=-2.8+0.75 \log M$ with a uniform
dispersion of 0.2 around this correlation; this is
based on the trend observed for  Galactic clusters located at
large galactocentric distances which are 
those more likely to keep memory of the initial conditions. In
deriving this relationship we have considered the possibility that
some of the  clusters with the highest concentrations could have  initially
lower concentrations and so we have based our choice on the properties
of clusters with lower concentrations. The
 spread around the adopted $c-\log M$ relationship is adopted because
 at least part of the observed broad dispersion in this relationship
is likely to be primordial (McLaughlin 2000). 

For the host galaxy  we have adopted an isothermal model with constant
circular velocity and we have studied the evolution of GCS for a 
sample of 295 host galaxies with  values of
the effective radius and of the effective mass equal to those
determined observationally by Burstein et al. (1997). 
We have focussed our attention 
on giant galaxies with $\log M_e>10.5$ for which good observational
constraints on the GCLF are available (see e.g. Kundu \& Whitmore
2001, Larsen et al. 2001).
\section{Results}
\begin{figure*}
\plotone{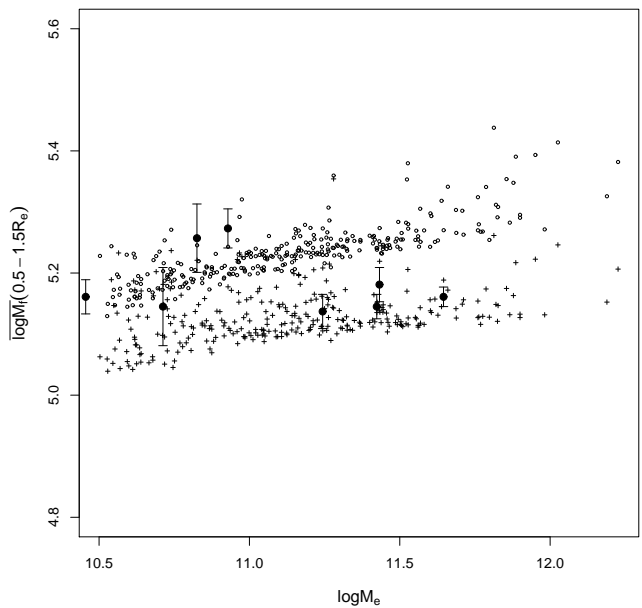}
\plotone{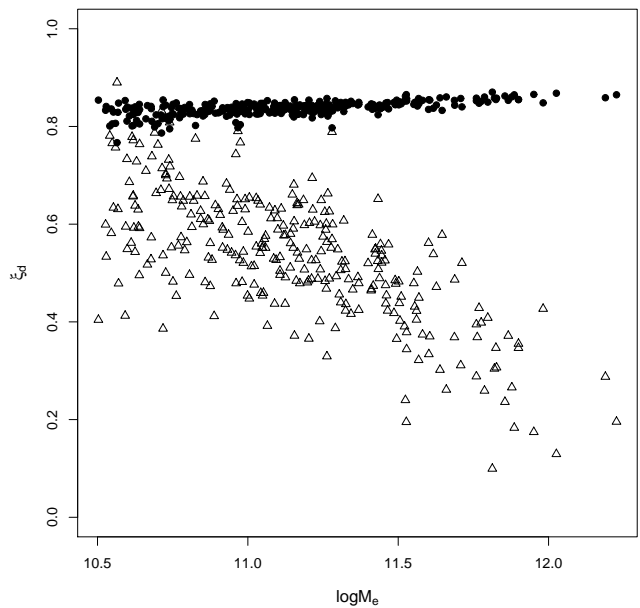}
\caption{(left panel) Final mean mass of clusters with projected
galactocentric distances between 0.5$R_e$ and 1.5 $R_e$ versus the
logarithm of the effective mass of the host galaxy. Crosses (open
dots)  show the final mean masses from  
simulations with $m_T=0.5~m_{\odot}$ ($m_T=0.9~m_{\odot}$) (see section
2 for the definition of $m_T$). Filled dots show the observed GCS mean
masses for a sample of giant elliptical galaxies from Kundu \& Withmore
(2001) ($M/L_V=2$ has been adopted to convert luminosities to masses).}
\caption{(right panel) Fraction of the initial
population of clusters 
disrupted because of mass loss due to stellar evolution (filled dots)
as a function of the 
effective mass of the host galaxy. Triangles show the fraction of
clusters unscathed by dissolution due to mass loss through stellar
evolution which are subsequently disrupted by dynamical friction and
two-body relaxation.}  
\end{figure*}

In Figure 1 we plot the final GCS mean masses from our simulations
versus the effective mass of the host galaxies. The values of the mean
masses plotted are those of clusters with projected galactocentric distances
between $0.5 R_e$ and $1.5 R_e$ which is the range of galactocentric
distances covered by observational data.
 The observational data shown in figure 1 are from
the analysis of Kundu \& Whitmore (2001) ($M/L_V=2$ has been adopted
to convert luminosities to masses).

As discussed in the Introduction above, in  previous studies of the
evolution of GCSs  not including  the effect of disruption of 
low-concentration clusters, it was found that for 
GCSs starting with a power-law initial GCMF similar to that observed in young
cluster systems, the final GCS mean masses were smaller than those
observed, the galaxy-to-galaxy variation of $\overline{\log M}_f$ and 
its radial
variation within individual galaxies were too large (see Vesperini
2001, Vesperini et al. 2002 and references therein).
The results of these new simulations including  disruption of
low-concentration clusters induced by  mass loss associated with stellar
evolution clearly show that this additional process plays a
significant role in determining the final values of the GCS mean masses.
Both the final values of the mean masses and  the galaxy-to-galaxy
variation of $\overline{\log M}_f$ are consistent with observational
data.
For the large majority of the host galaxies considered, the final mean mass
does not vary significantly with galactocentric distance: the average
difference between the mean mass of inner 
clusters (defined here as those with galactocentric distance smaller
than $R_e$) and outer clusters (those with
galactocentric distance larger than $R_e$) is 
 $\langle \Delta \overline{\log M}_f \rangle \simeq -0.06$; this value
is  smaller than that found in simulations not including
low-concentration cluster disruption  (see e.g. Vesperini 2001) and
consistent with observations.  

Also shown in figure 1 are the values of $\overline{\log M}_f$ from 
simulations starting with a different turnover in the stellar IMF
($m_T=0.9~m_{\odot}$); for clusters with a given mass and size, the
critical value of 
the concentration below which mass loss due to stellar evolution leads
to cluster dissolution increases with the fraction 
of massive stars (see e.g. Chernoff \& Shapiro 1987); as a
consequence of that and of 
the initial correlation between mass and concentration adopted in our
models, the values of $\overline{\log M}_f$ obtained in  simulations 
with $m_T=0.9~m_{\odot}$ are slightly larger than those in simulations with
$m_T=0.5~ m_{\odot}$. 

Although dynamical friction and mass loss due to two-body relaxation
lead to the disruption of a significant fraction of
clusters, the process of dissolution of low-concentration
clusters considered in this investigation appear to be required to
obtain final GCMF properties consistent with observations from a
power-law initial GCMF.  

As discussed in detail in Vesperini (1998) and in  
Vesperini (2000), when relaxation and dynamical friction act on a
bell-shaped GCMF with a mean mass similar to that of old GCSs (which
in this case is produced by disruption of low-concentration clusters) 
their effects do not manifest themselves in a
significant evolution of the mean mass; nevertheless, these processes
are always important as they considerably reduce the number of
clusters in GCSs and change the properties of those which
survive. Note that  for the initial GCMF considered in this study, the
number of clusters massive enough to be affected by dynamical friction
is small and most of the GCMF evolution occurring after the disruption
of low-concentration clusters is determined by mass
loss and disruption due to relaxation.

Figure 2 illustrates the efficiency of different dynamical processes:
this figure shows the fraction of the initial population of clusters disrupted
as a result of the dissolution of low-concentration clusters and the
fraction of the population unscathed by this process which is
subsequently disrupted by evaporation due to
two-body relaxation and dynamical friction as a function of the
effective mass of the host 
galaxy. 

Dissolution of
low-concentration clusters induced by  mass  
loss due to stellar evolution is mostly determined by the initial
concentration of individual clusters and it depends only very
weakly on the mass of the host galaxy.
On the other hand, the efficiency of disruption caused by relaxation and
dynamical friction strongly 
depends on the properties of the host galaxies and the fraction of
clusters disrupted by these processes is smaller for more
massive galaxies. As already discussed
in Murali \& Weinberg (1997) and in Vesperini (2000), the dependency
of the fraction of surviving clusters on the host galaxy mass can give rise to
a trend between cluster specific frequency, defined as the
number of clusters per galaxy unit luminosity (Harris \& van den Bergh
1981), and the
mass or luminosity of the host galaxy consistent with  that suggested
by observations (see e.g. Ashman \& Zepf 1998, Elmegreen 2000 for reviews). 

We emphasize that the joint effect of all the dynamical processes
considered leads to the disruption of a significant fraction of the
initial population of clusters; according to the results of our
simulations, for the host galaxies considered in this study the
current number of clusters within the whole range of galactocentric
distances considered in this study is, depending on the properties of
the host galaxy, between 3 and 12 per 
cent of the initial population. Any investigation using estimates of
the current number of clusters in galaxies to study the efficiency of
cluster formation should not leave out of consideration the effects of
dynamical evolution and the dependence of their efficiency on the
properties of the host galaxy and on the galactocentric distance.
As to the mass of clusters, the 
current total mass in clusters located within the range of
galactocentric distances considered here is
between 10 and 43 per cent of the total initial mass of clusters
within the same range of galactocentric distances. 

\section{Conclusions}
In this paper we have studied the role of the dissolution of  
low-concentration clusters induced by mass loss due to stellar
evolution on the properties of GCSs in elliptical galaxies.

We have found that the dissolution of  
low-concentration clusters can significantly affect the GCMF
evolution; in particular we have shown that this process can transform
a power-law initial GCMF   
similar to that observed in young cluster systems into a bell-shaped
GCMF with a mean mass similar to that currently observed in old GCSs
before the effects of two-body relaxation and dynamical friction 
become dominant. Relaxation and dynamical friction then lead to
the additional disruption of a significant fraction of the remaining
clusters with relaxation playing a dominant role; 
however, as shown in Vesperini (2000),  
when two-body relaxation and dynamical friction act on a bell-shaped
GCMF with parameters similar to those 
observed in old GCSs, the final values of the GCS mean mass and the
galaxy-to-galaxy variation of the GCS mean mass are perfectly consistent with
observations. 

Dissolution of low-concentration clusters therefore may provide the
missing link between the power-law GCMF observed in young cluster
systems in merging galaxies and the bell-shaped GCMF required to
obtain, after the additional evolution due to two-body relaxation and
dynamical friction, final
GCMF properties consistent with observations.

In a future work we will study the dependence of the evolution of the
GCMF  on the initial distribution of cluster
concentrations. We will also investigate the time evolution of the GCMF
and compare our results to observations of the GCMF in 
young and intermediate-age GCSs.
\section*{ACKNOWLEDGMENTS}
The authors gratefully acknowledge support from NASA via the ATP grant
NAG5-11320. We thank Keith Ashman and Arunav Kundu 
for many useful discussions and an anonymous referee for useful comments.
\section*{REFERENCES}
Ashman K.M., \& Zepf S.E., 1998, Globular Cluster Systems, Cambridge
University Press \\
Ashman K.M., Zepf S.E., 2001, AJ, 122, 1888\\
Baumgardt, H., 1998, A\&A, 330, 480\\
Baumgardt, H., \& Makino, J., 2003, MNRAS, in press\\
Burstein,D.; Bender, R., Faber, S., \& Nolthenius, R., 1997, AJ, 114, 1365\\
Chernoff, D.F. \& Djorgovski, S. 1989, ApJ, 339, 904\\
Chernoff,D.F., Kochanek, C.S., \& Shapiro S.L., 1986, ApJ, 309, 183\\ 
Chernoff, D.F., \& Shapiro, S.L., 1987, ApJ, 322, 113\\
Chernoff D.F.,\& Weinberg M.D., 1990, ApJ, 351, 121\\
Djorgovski, S., \& Meylan, G.,  1994, AJ, 108, 1292\\
Elmegreen, B.G., 2000, in Block D.L., Puerari I., Stockton A.,
Ferreira D., eds., Toward a New Millennium in Galaxy Morphology,p.469,
Kluwer, Dordrecht\\ 
Elmegreen, B.G., 2002, ApJ, 564, 773\\
Elmegreen, B.G., \& Efremov, Y.N., 1997, ApJ, 480, 235\\
Fall, S.M., Zhang, Q., 2001, ApJ, 561, 751\\
Fukushige T., \& Heggie, D.C., 1995, MNRAS, 276, 206\\
Harris, W.E., 2001, in Star
Clusters, Lectures for 1998 Saas-Fee Advanced School, L. Labhardt, B. Binggeli eds., (Springer-Verlag, Berlin, 2001)\\
Harris, W.E., \& van den Bergh, S., 1981, AJ, 86, 1627\\
Harris, W.E., \& Pudritz, R.E. 1994, ApJ, 429, 177\\
King, I.R., 1966, AJ, 71, 64\\ 
Kroupa, P., 2001, MNRAS, 322, 231\\
Kundu, A. \& Whitmore, B.C.,  2001, AJ, 121, 2950\\
Kundu, A., Zepf, S.E., \& Ashman, K.M. 2003, in prep.\\
Larsen, S., Brodie, J.P., Hucra, J.P., Forbes, D.A., Grillmair, C.J.,
2001, AJ, 121, 2974\\
McLaughlin, D.E., 2000, ApJ, 539, 618\\
Meylan, G. \& Heggie, D.C., 1997, A\&AR, 8, 1\\
Murali, C. \& Weinberg, M.D., 1997, MNRAS, 288, 767\\
Navarro, J., Frenk, C., \& White, S., 1996, ApJ, 462, 563\\
Okazaki, T., \& Tosa, M., 1995, MNRAS, 274, 48\\
Prata, S.W., 1971, AJ, 76, 1029\\ 
Takahashi, K., Portegies Zwart, S.F., 2000, ApJ, 535, 759\\
Vesperini, E., 1997, MNRAS, 287, 915\\
Vesperini, E., 1998, MNRAS, 299, 1019\\
Vesperini, E., 2000, MNRAS, 318, 841\\
Vesperini, E., 2001, MNRAS, 322, 247\\
Vesperini, E., Heggie, D.C., 1997, MNRAS, 289, 898\\
Vesperini, E., Zepf S.E., Kundu A., \& Ashman K.M., 2002, ApJ, in press\\
Whitmore, B.C.,  2002, in STScI Symp. 14, A decade of HST Science,
M.Livio, K.Noll, M. Stiavelli, eds., in press\\  
 \end{document}